\documentclass[aps,prl,twocolumn,amsmath,amssymb,groupedaddress]{revtex4-1}
\usepackage{epstopdf}
\usepackage{epsfig}
\usepackage{color}
\newcommand{\AF}[1]{\textcolor{red}{#1}}
\newcommand{\FSR}[1]{\textcolor{black}{#1}}
\newcommand{\strike}[1]{\textcolor{black}{}}

\usepackage{graphics}
\usepackage{bm}
\usepackage{amssymb,amsmath}
\usepackage{graphicx}
\usepackage[caption=false]{subfig}

\newcommand{\ket}[1]{\left|#1\right\rangle}

\begin{document}


\title{Controlled generation of higher-order Poincar\'{e} sphere beams from a laser}



\author{Darryl \surname{Naidoo}$^{1}$}
\author{Filippus \FSR{S.} \surname{Roux}$^{1}$}
\author{Angela \surname{Dudley}$^{1}$}
\author{Igor \surname{Litvin}$^{1}$}
\author{Bruno \surname{Piccirillo}$^{2}$}
\author{Lorenzo \surname{Marucci}$^{2,3}$}
\author{Andrew \surname{Forbes}$^{1,4}$}

\affiliation{$^{1}$CSIR National Laser Centre, P.O. Box 395, Pretoria 0001, South Africa}
\affiliation{$^{2}$Dipartimento di Fisica, Universit\`{a} di Napoli Federico II, Complesso Universitario di Monte Sant'Angelo, via Cintia, 80126 Napoli, Italy}
\affiliation{$^{3}$Consiglio Nazionale delle Ricerche (CNR)-SPIN, Complesso Universitario di Monte Sant'Angelo, via Cintia, 80126 Napoli, Italy}
\affiliation{$^{4}$School of Physics, University of the Witwatersrand, Private Bag 3, Johannesburg 2050, South Africa}
\affiliation{Email: aforbes1@csir.co.za}

\date{\today}

\begin{abstract}
The angular momentum state of light can be described by positions on a higher-order Poincar\'{e} (HOP) sphere, where superpositions of spin and orbital angular momentum states give rise to laser beams that have found many applications, including optical communication, quantum information processing, microscopy, optical trapping and tweezing and materials processing. Many techniques exist to create such beams but none to date allow their creation at the source. Here we report on a new class of laser that is able to generate all states on the HOP sphere. We exploit geometric phase control with a non-homogenous polarization optic and a wave-plate inside a laser cavity to map spin angular momentum (SAM) to orbital angular momentum (OAM). Rotation of these two elements provides the necessary degrees of freedom to traverse the entire HOP sphere. As a result, we are able to demonstrate that the OAM degeneracy of a standard laser cavity may be broken, producing pure OAM modes as the output, and that generalized vector vortex beams may be created from the same laser, for example, radially and azimuthally polarized laser beams. It is noteworthy that all other aspects of the laser cavity follow a standard design, facilitating easy implementation.
\end{abstract}

\pacs{\AF{}}

\maketitle

\section*{Introduction}

\begin{figure*}[htbp]
\centerline{\includegraphics[width=16cm]{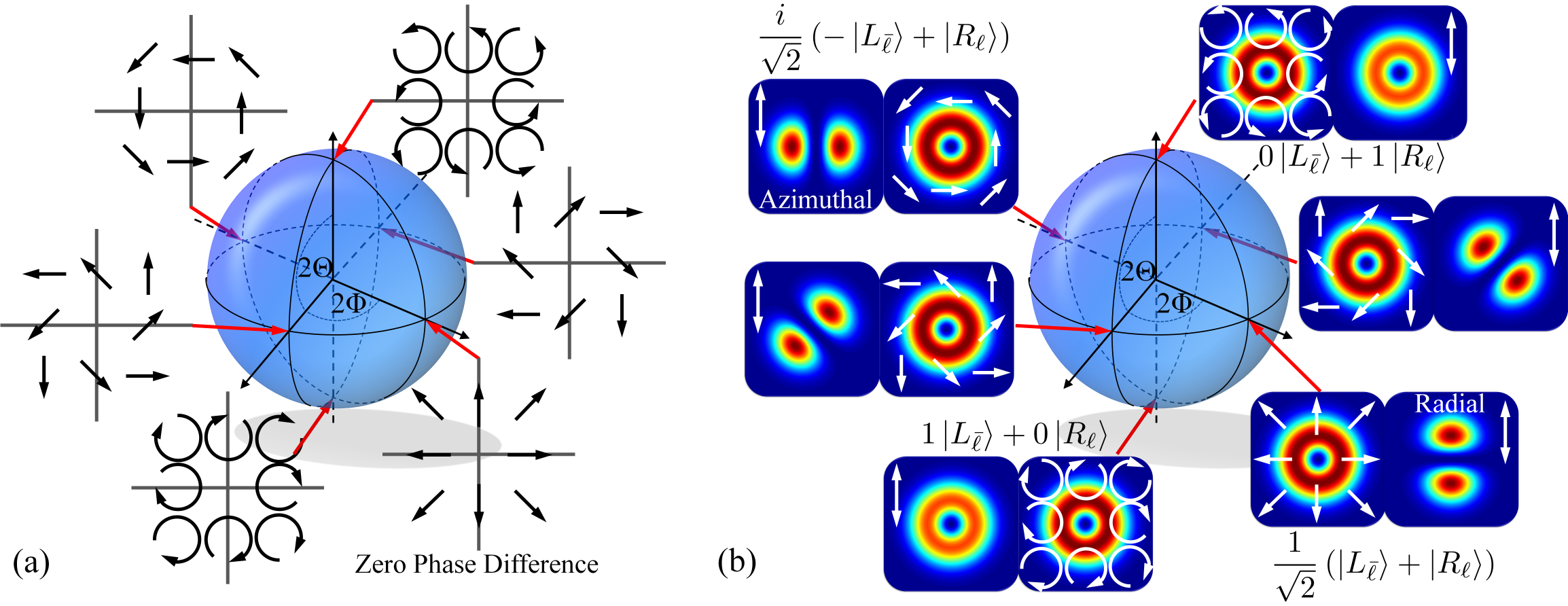}}
\caption{(color online). Higher-order Poincar\'{e} sphere representation of vector vortex beams illustrating the (a) local polarization vectors states at various positions on the sphere. The intensity of the outputs are (b) consistently beams with a central intensity null. These beams are differentiated by the transmitted intensity from a linear polarizer oriented in the vertical, as depicted by the double sided arrows. Expressions are provided for the states at the poles and for the special points on the equator with radial and azimuthal polarization.}
\label{fig:PSphere}
\end{figure*}

Recently the concept of the Higher-Order Poincar\'{e} (HOP) sphere was introduced as a theoretical framework for describing the total angular momentum of light, both spin and angular components \cite{Milione2011,Milione2012,Holleczek2011}. The HOP sphere describes higher-order states of polarization of generalized vector vortex beams, as shown in Fig.~\ref{fig:PSphere}~(a), in contrast to the Poincar\'{e} sphere (PS) which is a geometric representation of all possible states of polarization. While the Poincar\'{e} sphere is a Bloch sphere where the basis states are two orthogonal states of polarization, the HOP sphere is a Bloch sphere where the basis states are more general orthogonal states that incorporate both SAM and OAM. 

All the optical modes on the HOP sphere have an intensity distribution with a central null, as shown in Fig.~\ref{fig:PSphere}~(b). These states may be differentiated by the transmitted intensity through a linear polarizer, e.g., vertically orientated as depicted by the double sided arrows.  Light fields described by points on the HOP sphere are prevalent in nature and have found applications in high-speed kinematic sensing \cite{Berg2015}, OAM fiber mode selection \cite{Gregg2015}, space division multiplexing \cite{Lavery2014} and mode division multiplexing \cite{Milione2015}. In particular it is worth calling the attention of the reader to the poles and the equator of the sphere. The equator represents the cylindrical vector (CV) beams \cite{Zhan2009}, with special cases being the azimuthally and radially polarized light fields as shown in Fig.\ref{fig:PSphere}~(b). These fields have found many applications, for example, in laser material processing \cite{Nesterov2000,Meier2007,Venkatakrishnan2012}, particle acceleration \cite{Tidwell1993,Gupta2007,Wong2010,Dai2011}, optical trapping \cite{Kozawa2010,Huang2012,Donato2012,Loke2014} and microscopy \cite{Hao2010,Chen2013}. The extra-cavity generation of such beams has been achieved by using an interference approach \cite{Tidwell1990,Tidwell1993,Passilly2005}, liquid crystals \cite{Ren2006,Bashkansky2010}, sub-wavelength grating \cite{Machavariani2007} and from a spirally varying retarder \cite{Lai2008}. Laser cavities have been customized to produce particular CV beams by techniques such as inducing thermal stress to isotropic gain media \cite{Moshe2003}, by exploiting the thermal birefringence of laser gain media \cite{Yonezawa2006,Kawauchi2008,Ito2009}, with the use of an intra-cavity axicon \cite{Kozawa2005,Bisson2006,Chang2013} and with a conical shaped pump beam \cite{Wei2013,Vyas2014,Fang2015}.
The poles of the HOP sphere represent scalar vortex beams (having helical wavefronts) with a uniform circular polarization (right circular at the north pole and left circular at the south pole). The helicity of the wavefront arises from the azimuthally varying phase structure of $\exp(i \ell \phi)$ and such beams carry orbital angular momentum (OAM) of $\ell \hbar$ per photon where $\ell$ is referred to as the topological charge and can take any integer value. Henceforth we will refer to the sign of the helicity (the sign of $\ell$) as the ``handedness'' of the light. Such beams have found many applications in diverse fields such as optical manipulation \cite{Grier2003,Padgett2011}, and optical free space communication \cite{Wang2012}. While many attempts have been made to generate these modes inside a laser cavity \cite{Senatsky2012,Lin2014,Kim2015,Lin2015,Litvin2014} the degeneracy in the handedness of the azimuthal modes means that standard laser cavities cannot distinguish them: the spatial intensity distribution of laser modes with opposite azimuthal handedness (such as $+\ell$ and $-\ell$) are identical, they have identical radii of curvature on the wavefront and identical Gouy phase shifts. Consequently their intra-cavity losses are identical and thus very often uncontrolled helicities, coherent, or incoherent superpositions of modes with opposite handedness are produced \cite{Litvin2014}. Thus while customized lasers have demonstrated specific points on the HOP sphere, each point requiring its own laser design, no laser to date has been able to create an arbitrary HOP sphere beam.  

Here we show the generation of any HOP sphere beam directly from a laser. We couple SAM to OAM inside the laser cavity by means of a wave-plate and a non-homogeneous polarisation optic ($q$-plate) so that polarization control maps to OAM mode control. This is the first time that Pancharatnam-Berry (geometric) phase control has been applied inside a laser for mode selection. By control of the relative angles between the wave-plate and $q$-plate we can adjust the geometric phase change of the circulating light, and use this to produce any arbitrary beam on the HOP sphere, including the special cases of cylidrical vector vortex beams, e.g., azimuthally and radially polarised light, as well as pure OAM modes. We outline the theory for the mode control and confirm it experimentally in a solid state laser for HOP sphere beams of azimuthal orders $|\ell| = 1$ and $|\ell| = 10$.

\section*{Concept and Theory}
In contrast to the complexity and challenges of producing OAM beams and vector vortex beams from lasers, the control of polarization, or spin angular momentum (SAM), inside laser cavities is a well established technique \cite{Hodgson2005}. Our central idea is to exploit the SAM control as a proxy for OAM control, thereby realising generalized modes on the HOP sphere. 

\begin{figure}[htbp]
\centerline{\includegraphics[width=8.3cm]{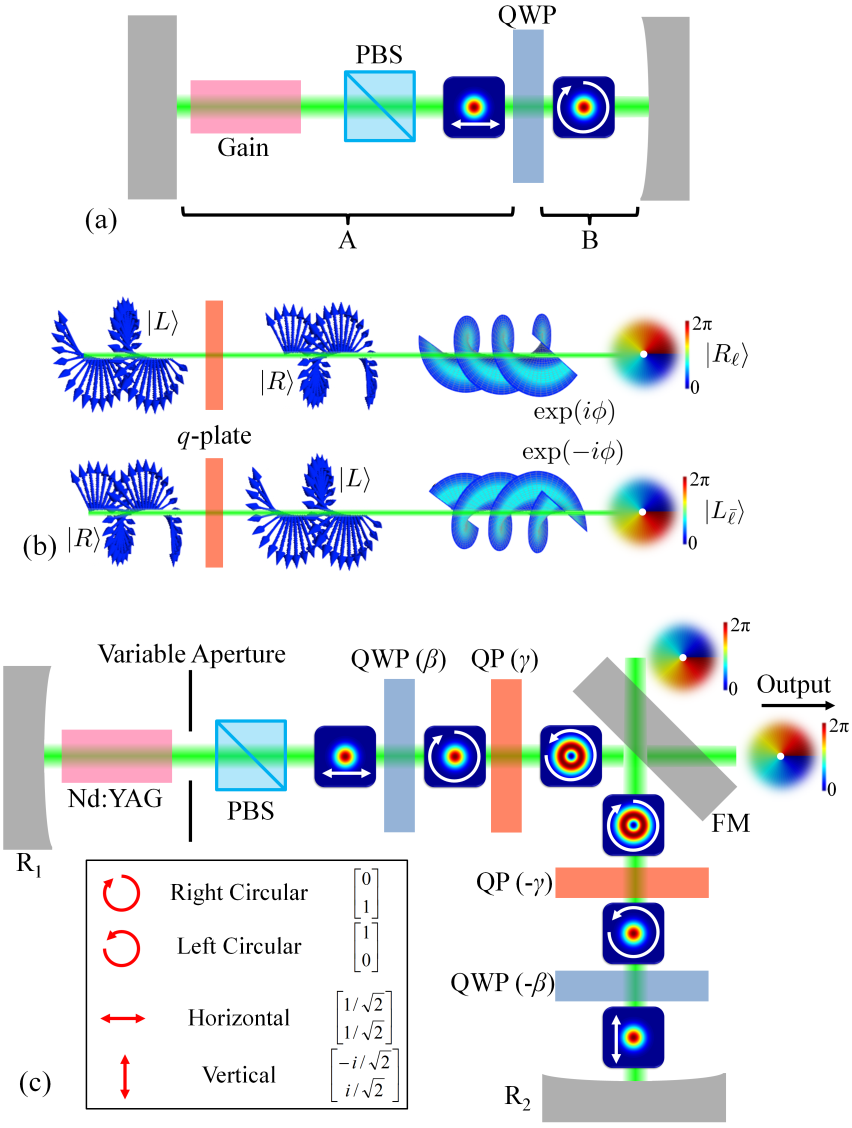}}
\caption{(color online). (a) The selection of linearly polarized light in a standard solid-state laser cavity in a Fabry-P\'{e}rot configuration is performed with a polarizing beam splitter (PBS) and the use of a quarterwave-plate (QWP) controls the fraction of light leaked out. (b) The $q$-plate is used to transform some single mode light beam into a helically-phased beam depending on the incident state of circular polarization and the handedness of the output beam is achieved through the following selection rules. (c) Experimental concept of the active selection pure OAM LG$_{0 \ell}$ modes of opposite handedness by the intra-cavity coupling of SAM to OAM. The coupling is achieved by selecting a pure SAM state by transmitting light that is linearly polarized in the horizontal through a quarterwave-plate (QWP) rotated at some angle $\beta$. This LG$_{00}$ shaped field is directed to a $q$-plate (QP) rotated at some angle $\gamma$ and consequently out coupled through the folding mirror (FM). The two rotation angles may be varied accordingly to map out the higher-order Poincar\'{e} sphere. The inset illustrates the various polarization states operating in the cavity with their associated vectors described in the circular polarization basis.}
\label{fig:setup}
\end{figure}

Consider a standard solid-state laser cavity in a Fabry-P\'{e}rot configuration, as shown in Fig.~\ref{fig:setup}~(a). Inclusion of a polarising beam splitter (PBS) and quarterwave-plate (QWP) ensure that the polarisation state in region A is always linearly polarised, the orientation dependent on that of the PBS. Traditionally such cavities are used to output light from the PBS, with the orientation of the QWP acting as a control on the fraction of light leaked out. It follows that in region B the circulating light is circularly polarised. In such a cavity the polarisation at any position is controlled and repeated after every round trip. 
We introduce a non-homogenous polarisation optic, in the form of a $q$-plate \cite{Marucci2006}, into the cavity to act as a SAM to OAM converter. The $q$-plate ladders some incoming OAM state following the selection rules: 
$\left|\ell,L\right\rangle \rightarrow \left|\ell+2q,R\right\rangle$ and $\left|\ell,R\right\rangle \rightarrow \left|\ell-2q,L\right\rangle$, where $L$ and $R$ refer to left and right circularly polarised light, $\ell$ is the incoming OAM state and $q$ is the charge of the $q$-plate (see Supplementary Material). 
This concept is illustrated graphically in Fig.~\ref{fig:setup} (b). By modifying the standard cavity to that shown in Fig.~\ref{fig:setup} (c), OAM-carrying beams are created within the cavity. The doubling of the elements ensures that the spatial mode and polarisation states are repeated after each complete round trip. The QWP and $q$-plate angles provides two degrees of freedom necessary to traverse the entire HOP sphere. It can be shown (see Supplementary Material) that our repeating mode in the cavity can be described by 
\begin{equation}
\textbf{v}_{out} = \left[ \cos\left(\frac{\Theta}{2}\right) \exp\left(-i\frac{\Phi}{2}\right) \right] \left|L_{\bar{\ell}}\right\rangle + \left[\sin\left(\frac{\Theta}{2}\right) \exp\left(i\frac{\Phi}{2}\right)\right] \left|R_{\ell}\right\rangle ,
\label{Eq4}
\end{equation}
where $\left|L_{\bar{\ell}}\right\rangle = \exp(-i |\ell| \phi) \left|L\right\rangle$, $\left|R_{\ell}\right\rangle = \exp(i |\ell| \phi) \left|R\right\rangle$, with $|L\rangle$ and $|R\rangle$ representing uniform left circular and right circular polarization states, respectively, $\Theta=\pi/2+2\beta$ and $\Phi=2\gamma-2\beta$ where $\beta$ and $\gamma$ are the rotation angles of the QWP and q-plate, respectively. This is precisely the description of a point on the HOP sphere with coordinates $\Theta$ and $\Phi$, where the poles on the sphere represent the basis states $\left|R_{\ell}\right\rangle$ and $\left|L_{\bar{\ell}}\right\rangle$. In other words, any HOP sphere beam can be realised from the laser. Examples of special cases are given in Table \ref{Table1}. 

\begin{table} [htbp]
\caption{Specialised output states are realised by rotating the QWP and $q$-plate through angles $\beta$ and $\gamma$, respectively. These include a pure OAM state of $-\ell$ ($-\pi/4,\gamma$), pure OAM state of $+\ell$ ($\pi/4,\gamma$), radial polarization ($0,0$) and azimuthal polarization ($0,\pi/2$) }
\centering
\setlength{\tabcolsep}{10pt}
\renewcommand{\arraystretch}{1.7}
\begin{tabular}{ |c||c|c| } 
\hline
$\beta$ & $\gamma = 0$ & $\gamma = \pi/2$ \\
\hline\hline
$-\pi/4$ & $e^{-i\frac{\pi}{4}}\left|L_{\bar{\ell}}\right\rangle+\left|R_{\ell}\right\rangle$ & $e^{-i\frac{3\pi}{4}}\left|L_{\bar{\ell}}\right\rangle+0\left|R_{\ell}\right\rangle$ \\ 
\hline
0 & $\frac{1}{\sqrt{2}}\left(\left|L_{\bar{\ell}}\right\rangle+\left|R_{\ell}\right\rangle\right)$ & $\frac{i}{\sqrt{2}}\left(-\left|L_{\bar{\ell}}\right\rangle+\left|R_{\ell}\right\rangle\right)$ \\ 
\hline
$\pi/4$ & $0\left|L_{\bar{\ell}}\right\rangle+e^{-i\frac{\pi}{4}}\left|R_{\ell}\right\rangle$ & $0\left|L_{\bar{\ell}}\right\rangle+e^{i\frac{\pi}{4}}\left|R_{\ell}\right\rangle$ \\
\hline
\end{tabular}
\label{Table1}
\end{table}

Heuristically the cavity can be understood by following the evolution of a Gaussian mode of linear polarisation propagating in region A away from mirror $R_1$. The horizontally polarised Gaussian beam is converted into a left circularly polarised Gaussian beam after the wave-plate if the wave-plate axis is at $45^o$. The $q$-plate converts this left circularly polarised beam into an OAM beam of charge $\ell = 1$ with right circular polarization. Reflection off the mirror inverts the entire state in both SAM and OAM, while the two remaining elements, orientated at opposite angles to the first two, reverse the process to create a vertically polarised Gaussian beam incident on mirror $R_2$. When this beam is propagated backwards through the cavity the modes invert again and return to mirror $R_1$ to the starting mode. The consequence is that the handedness of the light, as well as its vector nature, is completely defined by the angles of the QWP ($\beta$) and $q$-plate ($\gamma$). For example, if the QWP is rotated to produce linearly polarized light prior to the $q$-plate, then superpositions of left and right handed light with opposite OAM charges is produced - our general vector beams.

\section*{Experiment} 
The resonator concept as illustrated in Fig.~\ref{fig:setup}~(c) necessitates the use of a pair of $q$-plates and a pair of QWPs with a polarization insensitive $45^{\circ}$ mirror (FM) positioned between the $q$-plates. This cavity may be equivalently constructed by resorting to a V-shaped cavity where the two arms are separated within a few degrees with a planar mirror positioned at the apex of the V allowing for an off-axis design where only a single $q$-plate and QWP are required. The V-shaped cavity was experimentally realised in a diode-pumped solid-state laser where a 0.5 at.-$\%$ Nd-doped YAG rod (4 $\times$ 50 mm rod) was side pumped with a total input average pump power of $\sim 600$ W operating at 805 nm. The end mirrors were both concave high reflectors with curvatures $R_{1} = 400$ mm and $R_{2} = 500$ mm, respectively, with a planar mirror of 90$\%$ reflectivity positioned at the apex of the V. The separation distance between the two concave mirrors was 900 mm and the angle at which the two arms were separated by the plane mirror was in the order of 5$^{\circ}$. The $q$-plate ($q = 1/2$) was designed to operate most efficiently when positioned on-axis and it was thus positioned sufficiently adjacent to the plane mirror. The QWP (Multi-order operating at 1064 nm) was required to transmit both arms and was thus positioned to incorporate its clear aperture of $\sim 12$ mm. A lens of focal length $f = 400$ mm was inserted in the cavity to aid stability and to facilitate the clear aperture restriction imposed by the QWP. Finally a polarizing beam splitter (PBS) was preferred for the selection of linear polarization in the horizontal. A further practical consideration was required to be met in that with the pump arrangement, multimode operation was favoured which allows the existence of higher-order azimuthal and radial modes, we thus inserted a circular aperture with variable diameter such that the field incident on the $q$-plate was LG$_{00}$ in shape, i.e., only radial indices of $p = 0$ were allowed. The forward propagating wave with this arrangement was considered as the propagation from $R_1$ to $R_2$ with the back propagating wave acting in reverse. These two waves impinged on the planar mirror thus presenting two outputs; however, our interest lies in the output of the forward wave, as described by Eq. \ref{Eq4}.

\section*{Results}

\begin{figure}[htbp]
\centerline{\includegraphics[width=8.3cm]{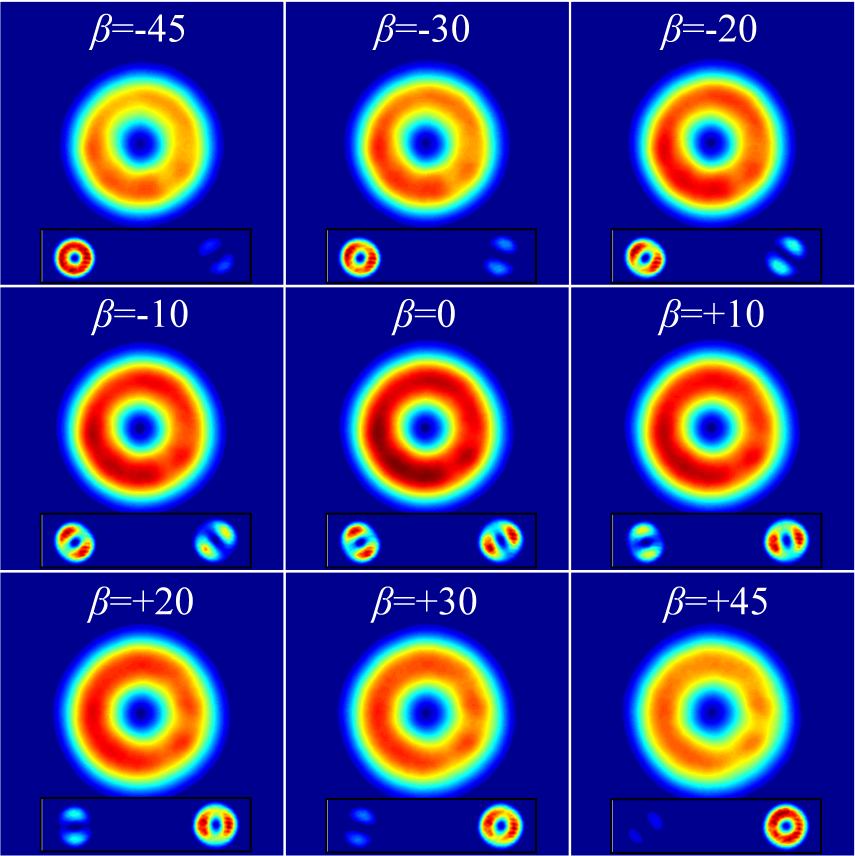}}
\caption{(color online). Annular shaped profiles are recorded at the output of the laser for the forward propagating wave for a variety of rotation angles of the QWP. The insets, showing filtered left- and right-circularly polarized components of the field, illustrate the state of polarization at the output where we evolve from a pure SAM state of left handedness ($\beta=-45^\circ$) to a pure SAM state of right handedness ($\beta=+45^\circ$) with an equivalent superposition of SAM states in between ($\beta=0^\circ$).}
\label{fig:Results1}
\end{figure}

We initially set $\gamma = 0$ and varied the angle of rotation, $\beta$, of the QWP. The output beam was an annular shaped beam (see Fig.~\ref{fig:Results1}) independent of $\beta$, as expected from theory. The state of the output is given by $\textbf{v}_{out} =\alpha_{1}\left|L_{\bar{\ell}}\right\rangle + \alpha_{2}\left|R_{\ell}\right\rangle$ where $\alpha_{1}$ and $\alpha_{2}$ are the relative weightings of the states on the poles. A measurement of the polarisation state (evident from the inserts in Fig.~\ref{fig:Results1}) confirms that the mode evolves from a left-circularly polarized beam ($\beta=-45^\circ$) to a right-circularly polarized beam ($\beta=+45^\circ$).

\begin{figure}[htbp]
\centerline{\includegraphics[width=8.3cm]{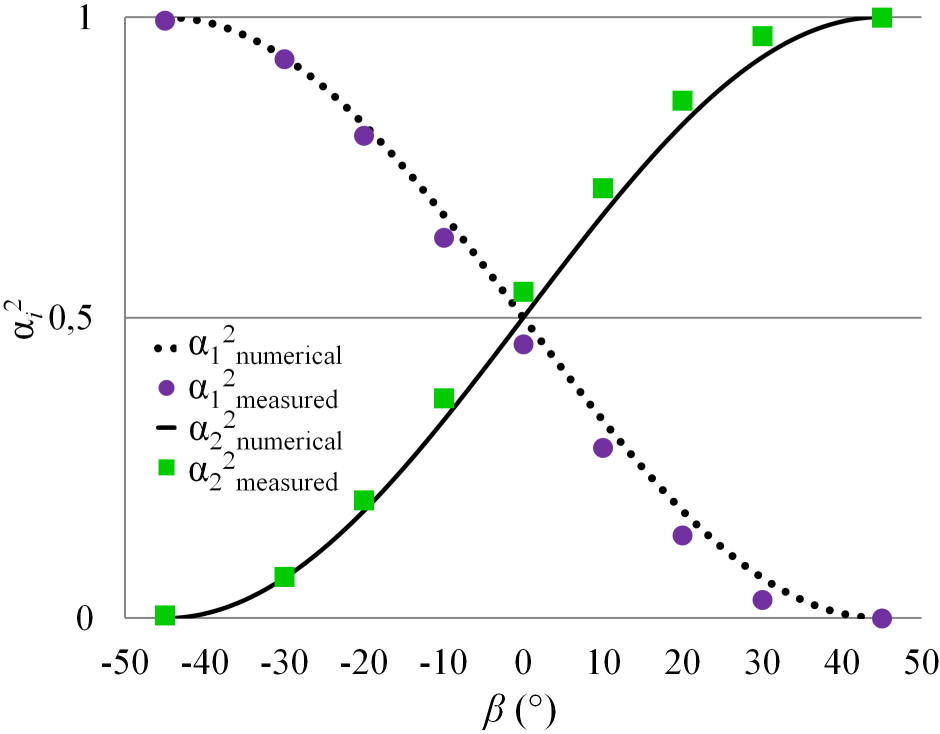}}
\caption{(color online). The variation in the polarization at the output is measured by determining the intensity of the relative weightings, $\alpha_{1}^{2}$ and $\alpha_{2}^{2}$, of the states on the poles, $\left|L_{\bar{\ell}}\right\rangle $ and $\left|R_{\ell}\right\rangle $, respectively. This compares well with the numerical determination for $\beta$ varied from $-45^\circ$ to $+45^\circ$. }
\label{fig:Results11}
\end{figure}

\begin{figure}[htbp]
\centerline{\includegraphics[width=8.3cm]{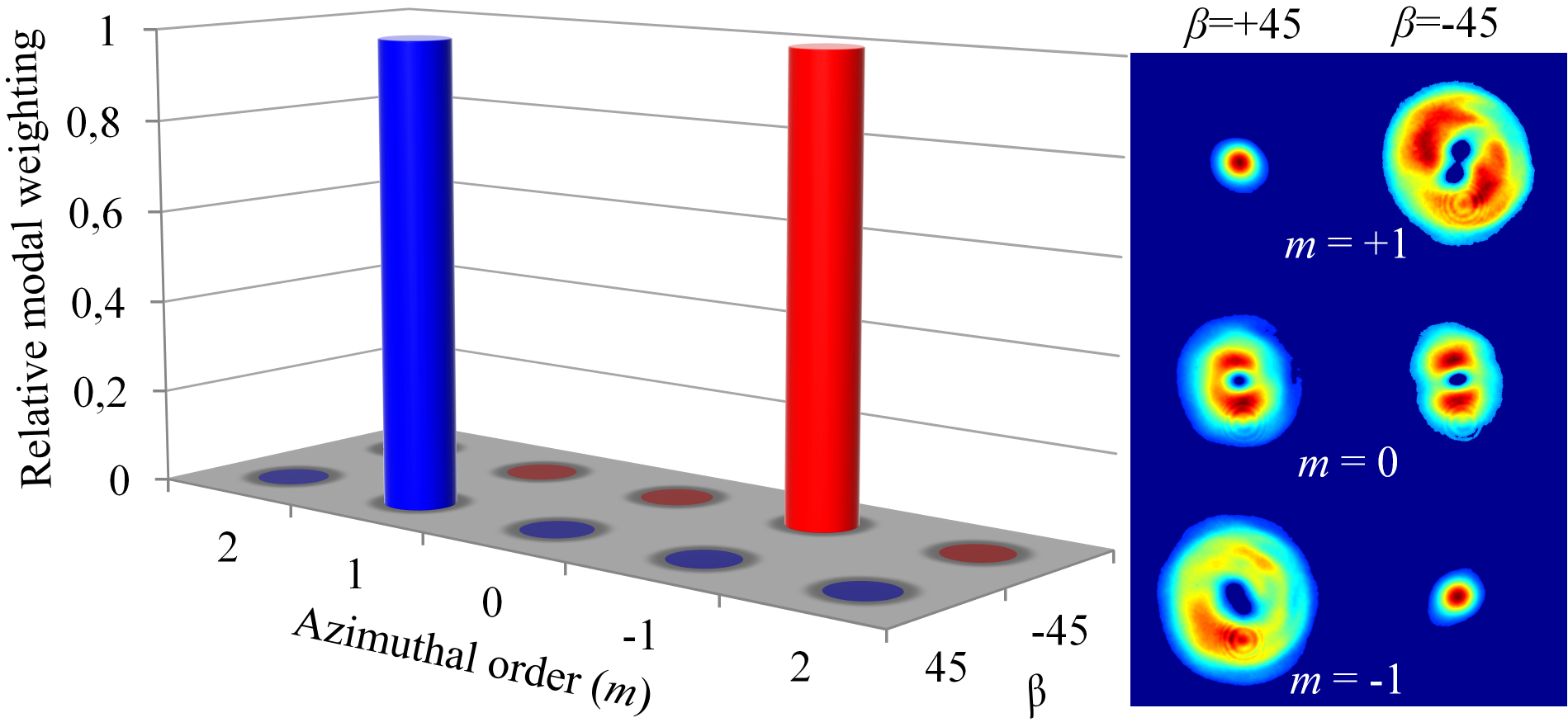}}
\caption{(color online). An azimuthal inner product is executed on the output of the laser operating under $\beta=-45^{\circ}$ and $+45^{\circ}$ illustrating a pure LG$_{0 -1}$ and LG$_{0 +1}$ mode, respectively, with their corresponding measurement channels.}
\label{fig:Results2}
\end{figure}

To determine the accuracy in the variation of the polarization as shown in Fig.~\ref{fig:Results1}, we measured the intensity of the relative weightings of the left and right components of the transmitted light. These components describe the states on the poles of the HOP sphere and thus $\alpha_{1}=\cos\left(\Theta/2\right) \exp\left(-i\Phi/2\right)$ and $\alpha_{2}=\sin\left(\Theta/2\right) \exp\left(i\Phi/2\right)$. The measured intensities of the respective components compare well with the numerical determination for $\beta$ varied from $-45^\circ$ to $+45^\circ$ as illustrated in Fig.~\ref{fig:Results11}.
Next we measured the OAM state by an azimuthal inner product \cite{Flamm2012,Naidoo2012} with a phase-only spatial light modulator (see Supplementary Material). We find that at $\beta = -45^{\circ}$ the mode is a pure $\ell = -1$ helicity, while a pure $\ell = +1$ for $\beta = +45^{\circ}$. This is illustrated graphically in Fig.~\ref{fig:Results2}, together with the raw data for three of the modal decomposition channels, where a central peak indicates the present mode, and a central null indicates the absence of that mode. The intra-cavity aperture ensures that the radial index of the mode is $p = 0$, and this too is confirmed by modal decomposition (see Supplementary Material). This approach presents a means to actively select the handedness of pure LG$_{0 \pm\ell}$ modes depending only on the rotation angle of the QWP $\beta$ and the charge $q$ of the $q$-plate.
Modes represented on the equator of a HOP sphere consist of a mixture of SAM and OAM states as determined by Eq.~(\ref{Eq4}). The combination of SAM states is achieved by setting $\beta$ to zero such that a pure linear state is incident on the $q$-plate resulting in a superposition output (as in the insert of Fig.~\ref{fig:Results1} for $\beta=0^\circ$). Consequently this also leads to a superposition of OAM and SAM states at the output, as given in Table \ref{Table1}. The non-separability of the polarization and spatial content of the mode means that upon passing through a linear polarizer, the annular shaped output splits into two lobes that rotate with a rotation in the polarizer. 

\begin{figure}[htbp]
\centerline{\includegraphics[width=8.3cm]{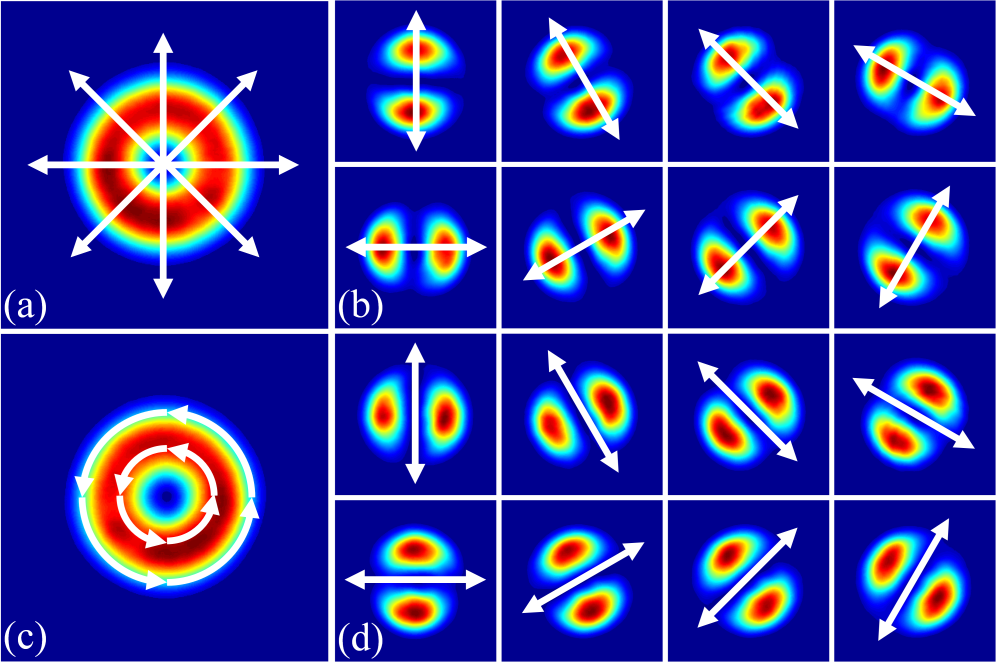}}
\caption{(color online). Pure vectorial vortex beams are actively selected by propagating linearly polarized light onto the $q$-plate by setting $\beta$ to zero. The (a) output of the laser at $\gamma$ set to zero is an annular shaped beam which is subsequently transmitted through a linear polarizer resulting in a lobed beam that angularly rotates with a rotation of the polarizer presenting a purely radially polarized beam. (b) The lobed structure is parallel to the orientation of the polarizer (indicated as double sided arrows) upon rotation. With $\gamma$ rotated by $90^\circ$ we select (c) an annular shaped profile that again delivers a lobed structure that rotates with a rotation in the linear polarizer, however, (d) perpendicularly to the orientation of the polarizer indicating pure azimuthal polarization.}
\label{fig:Results3}
\end{figure}

\begin{figure}[htbp]
\centerline{\includegraphics[width=8.3cm]{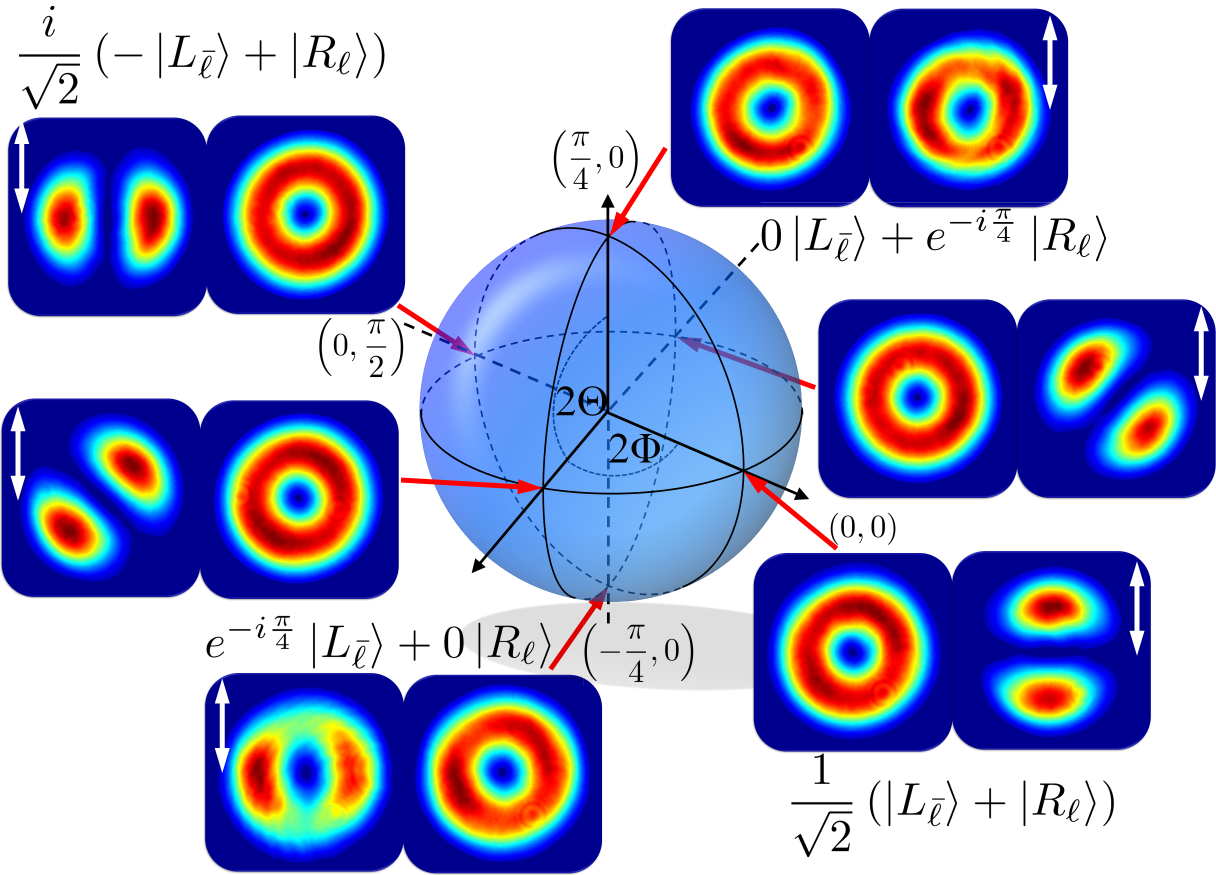}}
\caption{(color online). The experimental selection of beams as represented on a higher-order Poincar\'{e} sphere illustrating output beams that have an annular shaped intensity profile and are differentiated by the transmitted intensity from a linear polarizer oriented in the vertical as depicted by the double sided arrows. The states on the poles with the special cases of radial and azimuthal polarization are represented by the corresponding expressions where the values in parenthesis represent the rotation angles of $\beta$ and $\gamma$, respectively.}
\label{fig:Results4}
\end{figure}

With the laser operating under the conditions of $\beta = 0$ and $\gamma = 0$, we obtain an annular shaped (Fig.~\ref{fig:Results3}~(a)) beam that leads to a rotatable lobed beam (Fig.~\ref{fig:Results3}~(b)) succeeding a linear polarizer. The two lobed structure is oriented parallel to the orientation of the linear polarizer (illustrated as double sided arrows) thus presenting a pure radially polarized vectorial vortex beam. With $\gamma$ rotated by $90^\circ$ we select an annular shaped beam (Fig.~\ref{fig:Results3}~(c)) that is of pure azimuthal polarization which is hallmarked by the two lobed structure being perpendicular to the orientation of the linear polarizer (Fig.~\ref{fig:Results3}~(d)). The remarkable nature of selectively exciting these vectorial vortex beams is that not only are cylindrical vector-vortex beams achievable but so too are arbitrary vector states by controlling the input polarization state on the $q$-plate by adequately selecting the rotation angle $\beta$. An astute control of $\beta$ and $\gamma$ allows for the entire HOP sphere to be mapped and to aid consistency in comparison to Fig.~\ref{fig:PSphere}~(b), the states between the radial and azimuthal polarizations are accordingly determined as illustrated in Fig.~\ref{fig:Results4}. The annular outputs are transmitted through a linear polarizer oriented in the vertical (depicted by the double sided arrows) and are in excellent agreement with the anticipated intensities.

\begin{figure}[htbp]
\centerline{\includegraphics[width=8.3cm]{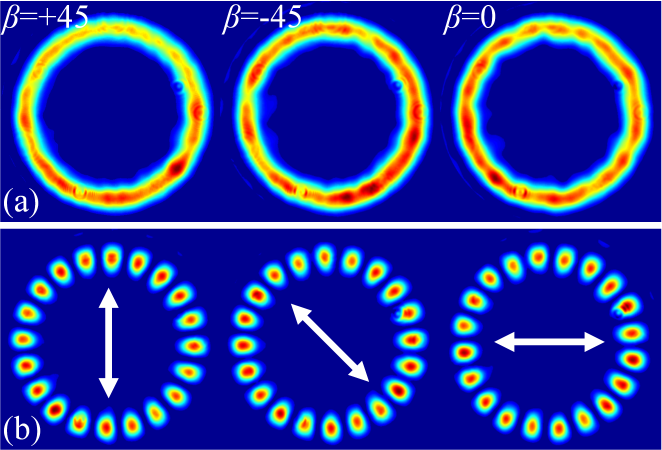}}
\caption{(color online). The technique in the selection of pure LG$_{0 \pm\ell}$ modes is not limited to $\ell=\pm 1$ and for a $q$-plate with $q=5$ we obtain (a) annular outputs for the cavity operating under $\beta=+45^\circ$, $-45^\circ$ and $0^\circ$ with $\gamma=0$. (b) At $\beta=0^\circ$, the output leads to a rotatable lobed beam succeeding a linear polarizer inferring radial polarization.}
\label{fig:Results5}
\end{figure}

\begin{figure}[htbp]
\centerline{\includegraphics[width=8.3cm]{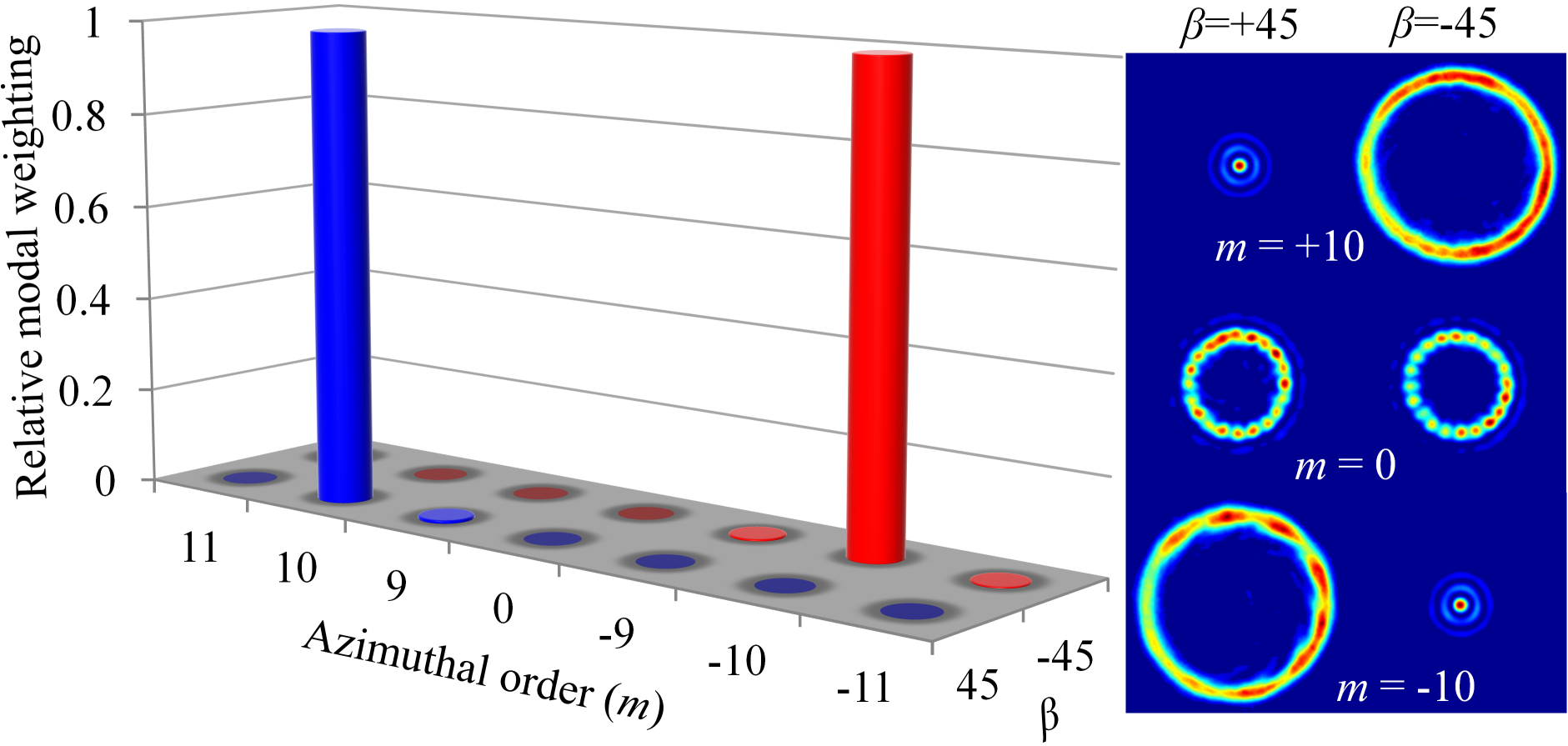}}
\caption{(color online). An azimuthal inner product is executed on the output of the laser operating under $\beta=-45^{\circ}$ and $+45^{\circ}$ illustrating a pure LG$_{0 -10}$ and LG$_{0 +10}$ mode, respectively, with their corresponding measurement channels.}
\label{fig:Results6}
\end{figure}

This technique is not limited to LG$_{0, \pm 1}$ modes, in fact a $q$-plate with a higher $q$ value may be equivalently realised. We demonstrate this by replacing the $q$-plate of $q=1/2$ with $q=5$ thus allowing for the selection of LG$_{0, \pm 10}$ modes without changing the physical properties of the cavity. The outputs for the cavity operating under $\beta=-45^\circ$, $+45^\circ$ and $0^\circ$ with $\gamma=0$ are illustrated in Fig.~\ref{fig:Results5}~(a) and show well defined annular beams. With the cavity operated at $\beta=0^\circ$, the annular output leads to a rotatable lobed beam succeeding a linear polarizer as shown in Fig.~\ref{fig:Results5}~(b) resulting in a radially polarized output. Again, to infer the OAM of the mode, we execute an azimuthal inner product with the digitally encoded transmission function $\exp ( i m \phi)$ for $m$ varying from -12 to +12 in unit steps and we identify an on-axis signal for $m = +10$ and $m = -10$ with zero elsewhere corresponding to operation for $\beta=+45^\circ$ and $-45^\circ$, respectively as presented in Fig.~\ref{fig:Results6} with some example measurement channels.

\section*{Conclusion}
We have outlined the concept for a new class of laser that utilised geometric phase control to realise arbitrary HOP sphere beams. We have demonstrated the concept in an otherwise conventional solid state laser cavity and shown the controlled generation of such beams, including the special cases of pure OAM modes as well as azimuthally and radially polarised light. As these fields have found many applications to date, we envisage that the versatility of creating HOP sphere beams directly from the source will find much interest. In particular, as the first example of intra-cavity mode selection by the Pancharatnam-Berry phase, we believe this report will spurn interest in this approach to designing custom lasers.




\newpage



\clearpage

\newpage
\section*{Supplementary Material}
\setcounter{equation}{0}
\subsection*{Quarterwave-plates and $q$-plates} 

The mechanism of operation of the laser is to a large extent determined by the operation of the quarterwave-plate and the $q$-plate, both of which are wave-plates. Here we provide a detailed discussion of the operation of these optical components, in terms of Jones matrices. For this purpose we'll use the circular polarization basis.

The Jones matrix for a general wave-plate (with either horizontal or vertical optic axis) is given by
\begin{equation}
U_{\rm WP} = \left[ \begin{array}{cc} \cos\left(\frac{\mu}{2}\right) & i\sin\left(\frac{\mu}{2}\right) \\ 
i\sin\left(\frac{\mu}{2}\right) & \cos\left(\frac{\mu}{2}\right) \end{array} \right] .
\label{wpmat}
\end{equation}
The Jones matrix for a rotated version of the general wave-plate is
\begin{eqnarray}
U_{\rm WP}(\beta) & = &  R(\beta) U_{\rm WP} R(-\beta) \nonumber \\
& = & \left[ \begin{array}{cc} \cos\left(\frac{\mu}{2}\right) & i e^{-i2\beta}\sin\left(\frac{\mu}{2}\right) \\ 
i e^{i2\beta}\sin\left(\frac{\mu}{2}\right) & \cos\left(\frac{\mu}{2}\right) \end{array} \right] ,
\label{wpmatr}
\end{eqnarray}
where the rotation matrix in the circular polarization basis is
\begin{equation}
R(\beta) = \left[ \begin{array}{cc} \exp(-i\beta) & 0 \\ 
0 & \exp(i\beta) \end{array} \right] .
\label{rmat}
\end{equation}

For a quarterwave-plate $\mu=\pi/2$, which gives

\begin{equation}
U_{\rm QWP}(\beta) =\frac{1}{\sqrt{2}} \left[ \begin{array}{cc} 1 & i e^{-i2\beta} \\ 
i e^{i2\beta} & 1 \end{array} \right] ,
\label{qwpmatr}
\end{equation}
and for a halfwave-plate $\mu=\pi$, giving
\begin{equation}
U_{\rm HWP}(\beta) = \left[ \begin{array}{cc} 0 & i e^{-i2\beta} \\ 
i e^{i2\beta} & 0 \end{array} \right] .
\label{hwpmatr}
\end{equation}

In the last five years, the use of $q$-plates in the generation of helicoidal beams has increased considerably \cite{Piccirillo2013}. These compact devices may ideally achieve conversion efficiencies (from a Gaussian to a helically phased mode) approaching 100\% and do not deflect the beam. By modulating the input polarization, they allow the selection of arbitrary states on the OAM Poincar\'{e} sphere with switching times in the order of a few nanoseconds. In addition, $q$-plates can be tuned for partial conversion or for the selection of the wavelength of the input beam and may be switched on and off by some external electrical control.

A $q$-plate is realized as a slab of a birefringent material, such as a liquid crystal, having a uniform birefringent phase retardation $\delta$ across the slab thickness (which can be electrically controlled) and a space-variant transverse optical axis distribution exhibiting a topological charge $q$ \cite{Piccirillo2010,Slussarenko2011}. The charge $q$ represents the number of rotations of the local optical axis in a path circling once around the center of the plate, where a topological defect must be present. The sign of $q$ may be positive or negative depending on whether the rotation of the axis has the same or opposite direction as the path.

In the simplified limit in which the $q$-plate is ideally thin, transverse diffraction effects arising from propagation inside the device can be neglected (such propagation effects have been discussed in Ref.~\cite{Karimi2009}, although only within an approximate treatment), so that the $q$-plate acts as an ideal phase optical element. In this approximation, the birefringence-induced Pancharatnam-Berry phase can be derived by using a simple Jones matrix approach \cite{Marucci2006,Marucci2006b,Marucci2008}. 
A $q$-plate is a halfwave-plate with an optic axis that varies as a function of the azimuthal angle. In other words, one needs to replace $\beta\rightarrow q\phi$ in Eq.~(\ref{hwpmatr}) \cite{note}, where $\phi$ is the azimuthal angle and $q$ is a half-integer (for continuity across the optical element). The Jones matrix for a general $q$-plate is therefore given by 
\begin{equation}
U_{\rm QP}(q) = \left[ \begin{array}{cc} 0 & i e^{-i2q\phi} \\ 
i e^{i2q\phi} & 0 \end{array} \right] .
\label{qpmat}
\end{equation}
We illustrate here the example for $q=1/2$ noting that other cases can simillarly be derived. Hence, discarding an overall factor of $i$, we obtain
\begin{equation}
U_{\rm QP} = \left[ \begin{array}{cc} 0 & e^{-i\phi} \\ 
e^{i\phi} & 0 \end{array} \right] .
\label{qpmat0}
\end{equation}

When operating on the circular polarization states, the $q$-plate in Eq.~(\ref{qpmat0}) produces
\begin{eqnarray}
U_{\rm QP} \ket{L} & = & \ket{R} e^{i\phi} \equiv \ket{R_{\ell}} \label{ldef} \\
U_{\rm QP} \ket{R} & = & \ket{L} e^{-i\phi} \equiv \ket{L_{\bar{\ell}}} .
\label{rdef}
\end{eqnarray}

The Jones matrix for the rotated version of the $q$-plate in Eq.~(\ref{qpmat0}) is
\begin{eqnarray}
U_{\rm QP}(\gamma) & = & \left. R(\gamma) U_{\rm QP} R(-\gamma) \right|_{\phi\rightarrow\phi-\gamma} \nonumber \\
& = & \left[ \begin{array}{cc} 0 & e^{-i(\phi+\gamma)} \\ 
e^{i(\phi+\gamma)} & 0 \end{array} \right] ,
\label{qpmatr}
\end{eqnarray}
where $\gamma$ is the q-plate rotation angle and, in deriving Eq.~(\ref{qpmatr}), in addition to the operation of the rotation matrices, which transform the polarization basis, one also needs to transform the coordinates, which leads to a shift in the azimuthal angle.

The combined operation of the quarterwave-plate and the $q$-plate, with rotation angles $\beta$ and $\gamma$, respectively, on a horizontally polarized input state, produces
\begin{eqnarray}
\ket{\psi}_{\rm out} & = & U_{\rm QP}(\gamma) U_{\rm QWP}(\beta) \ket{H}_{\rm in} \nonumber \\
& = & \left[ \begin{array}{c} \cos\left(\frac{\pi}{4}+\beta\right) e^{-i(\phi+\gamma-\beta)} \\ 
\sin\left(\frac{\pi}{4}+\beta\right) e^{i(\phi+\gamma-\beta)} \end{array}  \right] ,
\label{psidef}
\end{eqnarray}
where we discarded a global phase factor. The resulting output state can be expressed as
\begin{eqnarray}
\ket{\psi}_{\rm out} & = & \cos\left(\frac{\Theta}{2}\right) \exp\left(-i \frac{\Phi}{2}\right) \ket{L_{\bar{\ell}}} \nonumber \\
& & + \sin\left(\frac{\Theta}{2}\right) \exp\left(i \frac{\Phi}{2}\right) \ket{R_{\ell}} ,
\label{psiredef}
\end{eqnarray}
in terms of the basis states defined in Eqs.~(\ref{ldef}) and (\ref{rdef}), where $\Theta=2\beta+\pi/2$ and $\Phi=2\gamma-2\beta$. 

Note that Eq.~(\ref{psiredef}) represents an arbitrary position on the higher-order Poincar\'{e} sphere, denoted by coordinates $\Theta$ and $\Phi$, which are given in terms of the two physical rotation angles $\beta$ and $\gamma$.
\subsection*{Mode measurement}
 
An arbitrary function within a vector space may be represented as a linear combination of certain basis elements as there exists at least one basis set within the vector space. Similarly, an arbitrary paraxial optical beam may be expanded and represented by basis sets corresponding to orthogonal solutions to the paraxial wave equation. These solutions include Hermite-Gaussian, Ince-Gaussian and Laguerre-Gaussian functions and the linear combination based on these orthogonal sets may be expressed as:
\begin{equation}
U(\textbf{r}) = \sum\limits_{n=1}^{N}{a_{n}\Psi_{n}(\textbf{r})},
\label{eq:azField1}
\end{equation}
where $a_{n}=\rho_{n}e^{i\Delta\theta_{n}}$ is the complex correlation coefficient corresponding to a specific basis element $\Psi_{n}(\textbf{r})$ where $\Psi_{n}(\textbf{r})=\psi_{n}(\textbf{r})\textbf{e}_{n}$ is the $n^{th}$ mode having an amplitude, $\rho_{n}$, of a specific polarisation, $\textbf{e}_{n}$.  The phase difference between two modes $\Delta\theta_{n}$, is known as the intermodal phase difference where a mode with a planar phase is selected as one of the modes. The determination of the coefficients $a_{n}$ which are normalised according to:
\begin{equation}
\sum\limits_{n=1}^{N}\left|a_{n}\right|^{2}=\sum\limits_{n=1}^{N}\rho_{n}^{2}=1,
\label{eq:Field3}
\end{equation}
may be determined by executing an inner product given as:
\begin{equation}
a_{n}=\langle\textbf{U},\Psi\rangle=\iint_{\Re}U(\textbf{r})\Psi_{n}^{*}(\textbf{r})d^{2}\textbf{r},
\label{eq:azField2}
\end{equation}
where the asterisk represents the complex conjugate. The determination of the respective correlation coefficients allows for an arbitrary paraxial optical beam to be completely decomposed into the subsequent basis elements. The weightings are optically determined by sampling the resultant field $\left(u(x,y) = U(x,y)\Psi_{n}^{*}(x,y)\right)$ in the Fourier plane where the corresponding Fourier transformation is expressed as:
\begin{eqnarray}
U_{1}(k_{x},k_{y})=\mathcal{F}\left\lbrace u(x,y)\right\rbrace= \iint U(x,y)\Psi^{*}_{n}(x,y)\nonumber\\
\times\exp\left(-i\left(k_{x} x + k_{y} y\right)\right)dxdy.
\label{eq:azField4}
\end{eqnarray}
The weightings as expressed in Eq.~(\ref{eq:azField2}) are mathematically determined through an inner product and we obtain this formalism optically by measurement of the on-axis intensity of the field in the Fourier plane by setting the propagation vectors to zero ($k_{x}=k_{y}=0$) in Eq.~(\ref{eq:azField4}) and is expressed as:
\begin{eqnarray}
I^{\rho}_{n}(0,0)&=&\vert U_{1}(0,0)\vert^{2}\nonumber\\
&=&\left\vert \iint U(x,y)\Psi^{*}_{n}(x,y)dxdy\right\vert^{2}\nonumber\\
&=&\rho^{2}_{n}.
\label{eq:azField5}
\end{eqnarray}
In the decomposition of a field in both amplitude and phase, the type of transmission functions depends on the full field information of modes from an orthogonal set. For the extraction of an amplitude of a single mode from the orthogonal set, the transmission function may be chosen as the complex conjugate of the corresponding field:
\begin{equation}
T_{n}(\textbf{r}) = \psi_{n}^{*}(\textbf{r}).
\label{eq:Field4}
\end{equation}
With this transmission function the on-axis optical intensity in the Fourier plane (far-field) is $I_{n}^{\rho}\propto\rho_{n}^{2}$ which is in fact the power of mode $\psi_{n}$. The measurement of the intermodal phase difference of some mode $\psi_{n}$ to some reference mode $\psi_{0}$, however, requires two transmission functions where each represent an interferometric superposition of the two mode fields: 
\begin{eqnarray}
T_{n}^{\textrm{cos}}(\textbf{r}) = [\psi_{0}^{*}(\textbf{r})+\psi_{l}^{*}(\textbf{r})]/\sqrt{2}, \nonumber\\
T_{n}^{\textrm{sin}}(\textbf{r}) = [\psi_{0}^{*}(\textbf{r})+i\psi_{l}^{*}(\textbf{r})]/\sqrt{2}.
\label{eq:Field5}
\end{eqnarray}
The correlation of the incident field with these transmission functions results in intensities $I_{n}^{\textrm{cos}}\propto\rho_{0}^{2}+\rho_{n}^{2}+2\rho_{0}\rho_{n}\textrm{sin}\Delta\theta_{n}$ and $I_{n}^{\textrm{sin}}\propto\rho_{0}^{2}+\rho_{n}^{2}+2\rho_{0}\rho_{n}\textrm{cos}\Delta\theta_{n}$ corresponding to $T_{n}^{\textrm{cos}}$ and $T_{n}^{\textrm{sin}}$, respectively. Again, this is measured at the Fourier plane and the intermodal phase difference is calculated according to:
\begin{equation}
\Delta\theta_{n}=-\textrm{arctan}\left[\frac{2I_{n}^{\textrm{sin}}-\rho_{n}^{2}-\rho_{0}^{2}}{2I_{n}^{\textrm{cos}}-\rho_{n}^{2}-\rho_{0}^{2}}\right].
\label{eq:Field6}
\end{equation}
\begin{figure}[htbp]
\centerline{\includegraphics[width=8.3cm]{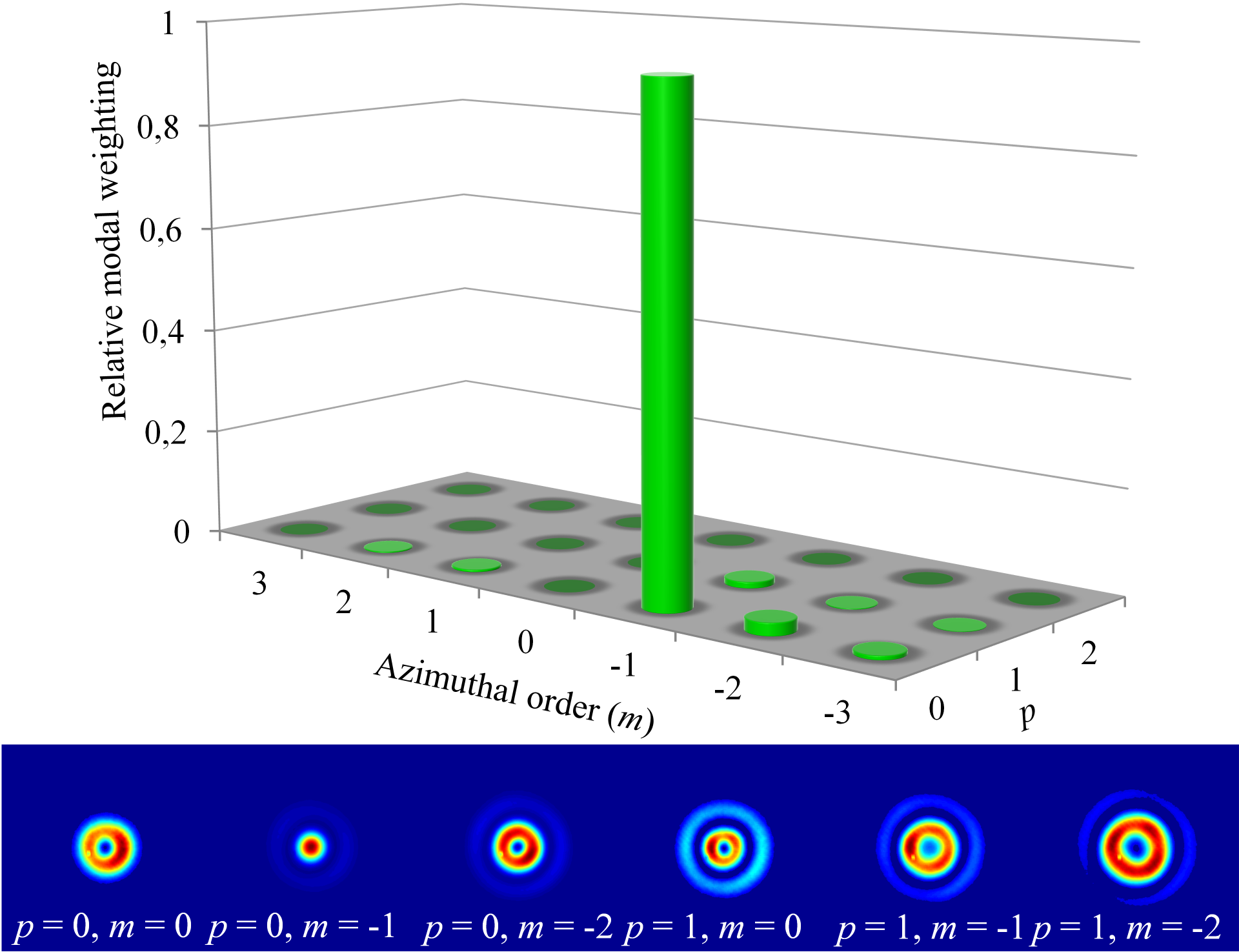}}
\caption{(color online). A modal decomposition in amplitude and phase was executed on the output of the cavity operated at $q=1/2$, $\beta=-45^\circ$ and $\gamma=0^\circ$. An intensity signal was obtained for a transmission function of LG$_{0 -1}$ with zero elsewhere which is further demonstrated in the measurement channels. }
\label{fig:Results7}
\end{figure}
Experimentally the unknown field $U(x,y)$ is directed onto a spatial light modulator (SLM) that is electronically addressed with an appropriate transmission function $T_{n}(\textbf{r})$. The resultant field is Fourier transformed with a thin optical lens by positioning the lens a focal length from the plane of the SLM and the intensity measurement to determine the relative weightings (Eq.~(\ref{eq:azField5})) is performed at a focal length beyond the lens. The transmission function as addressed to the SLM is coupled with a linear grating employed as a phase carrier to separate the first order of diffraction from the zeroth and unwanted diffraction orders and the intensity at the centre of the first diffraction order is sampled to determine the relative weightings. This experimental procedure was executed on the output of the optical cavity operated at $q=1/2$, $\beta=-45^\circ$ and $\gamma=0^\circ$. The basis set for the decomposition was chosen to be the Laguerre-Gaussian set and the transmission functions were varied from $p=0$ through $2$ with $\ell=-3$ through to $+3$ and as illustrated in Fig.~\ref{fig:Results7}, we obtain an on-axis intensity signal for a Laguerre-Gaussian mode of radial order, $p=0$ and azimuthal order $\ell=-1$ with zero for all other modes which affirms a pure LG$_{0 -1}$ mode at the output.  The Gaussian width in the Laguerre-Gaussian function was chosen from the cavity design parameters.

\begin{thebibliography}{1}
\bibitem{note}
For the special case $q=1$, one also needs to specify an additional angle such that $\beta\rightarrow \phi-\phi_0$. For the other cases the angle $\phi_0$ can be absorbed in the definition of the q-plate rotation angle $\gamma$, as defined in Eq.~(\ref{qpmatr}).
\end{thebibliography}

\end{document}